\documentclass[review]{elsarticle}
\usepackage{lineno,hyperref}
\usepackage{amsmath,amsthm,amssymb}
\usepackage{graphicx}
\usepackage{color}
\usepackage{dsfont}
\usepackage{epstopdf}
\usepackage[hang]{subfigure}
\usepackage{natbib}
\usepackage{textcomp}
\usepackage{algorithm}
\usepackage{algorithmicx}
\usepackage{algpseudocode}
\usepackage{bbm}
\usepackage{multirow}

\usepackage{ulem}

\newtheorem{thm}{Theorem}
\newtheorem{lem}{Lemma}

\newcounter{rmk}

\def\t{ \mathrm{\scriptscriptstyle T} }

\def\t{\mathrm{\scriptscriptstyle T} }

\def\tr{\text{tr}}
\def\ebic{\text{EBIC}}
\def\mx{\mathcal{X}}
\def\my{\mathcal{Y}}

\definecolor{red}{RGB}{139,0,18}
\definecolor{lightred}{RGB}{186,25,31}
\definecolor{blue}{RGB}{0,56,108}
\definecolor{lightblue}{RGB}{69,100,139}
\renewcommand\emph[1]{{\color{red}\itshape #1}}

\newcommand\red[1]{\textcolor{black}{#1}}

\journal{Knowledge-Based Systems}

\biboptions{authoryear}

\begin{document}
\begin{frontmatter}

\title{A sequential stepwise screening procedure for sparse recovery in high-dimensional multiresponse models with complex group structures}

\author[mymainaddress]{Weixiong Liang}
\author[mymainaddress]{Yuehan Yang\corref{mycorrespondingauthor}}
\cortext[mycorrespondingauthor]{Corresponding author}
\ead{yyh@cufe.edu.cn}

\address[mymainaddress]{School of Statistics and Mathematics, Central University of Finance and Economics}

\begin{abstract}
Multiresponse data with complex group structures in both responses and predictors arises in many fields, yet, due to the difficulty in identifying complex group structures, only a few methods have been studied on this problem. We propose a novel algorithm called sequential stepwise screening procedure (SeSS) for feature selection in high-dimensional multiresponse models with complex group structures. This algorithm encourages the grouping effect, where responses and predictors come from different groups, further, each response group is allowed to relate to multiple predictor groups. To obtain a correct model under the complex group structures, the proposed procedure first chooses the nonzero block and the nonzero row by the canonical correlation measure (CC) and then selects the nonzero entries by the extended Bayesian Information Criterion (EBIC). We show that this method is accurate in extremely sparse models and computationally attractive. The theoretical property of SeSS is established. We conduct simulation studies and consider a real example to compare its performances with existing methods.
\end{abstract}


\end{frontmatter}

% \linenumbers
\section{Introduction}
Feature selection in high-dimensional multiresponse models with complex group structures is important in many scientific fields, especially in genetic and medical studies. Many studies about detecting associations between multiple traits and predictors have found that both responses and predictors are often embedded in some biological functional groups, such as gene-gene associations \citep{park2008penalized,zhang2010new}, protein DNA associations \citep{zamdborg2009discovery}, and brain fMRI-DNA association study \citep{stein2010voxelwise,FAN2020106341}. The intrinsic group structures carry crucial information when modeling high-dimensional multiresponse models. However, these inherent and vital group structures are hard to distinguish.

Without considering the group structures, regular linear models can be classified into univariate response models and multiresponse models. Researchers have proposed many methods for the univariate response models. To select the true feature in univariate response models, one way is to use the penalized regularized regression approaches \citep{tibshirani1996regression, zou2006adaptive, zou2005regularization, fan2001variable, zhang2010nearly}; another efficient technique is to choose features at each step according to a specific criterion, which is called sequential procedures \citep{wang2009forward, ing2011stepwise, cai2011orthogonal, luo2014sequential}. Without considering group structures, these approaches can be extended to the multiresponse models by imposing specific constraints on the coefficient matrix $B$. There are various forms of constraints proposed from different aspects, such as from the singular value of $B$ \citep{yuan2007dimension}; from the $\text{rank}(B)$ \citep{chen2012sparse}; from the norms or the linear combination of norms of $B$ \citep{turlach2005simultaneous, obozinski2011support, peng2010regularized}.
\red{Recent years, many innovative methods for dealing with the linear models under specific situations have been proposed, such as regression with hidden variables \citep{bing2022adaptive}, low-rank regression \citep{cho2022multivariate}, estimation of the rank of the coefficient matrix \citep{bing2019adaptive}.}
These methods are efficient for the uncorrelated data, but they are not suitable for dealing with the data when both the univariate response models and multiresponse models share group structures \citep{biswas2012logistic,zhou2010association}.

With the increase of data dimensions, complex group structures are common in responses and predictors, seriously affecting the statistical modeling. Yet, due to the complexity, this kind of data has seldom been studied. As far as we know, there are a few researches focus on the high-dimensional multiresponse models with the group structures: 1) multivariate sparse group lasso (MSGL) \citep{li2015multivariate}, 2) sequential canonical correlation search (SCCS) \citep{luo2020feature}, 3) Bayesian linear regression \citep{ning2020bayesian}. MSGL is inspired from the group lasso \citep{yuan2006model} and sparse group lasso \citep{simon2013sparse}. \citet{li2015multivariate} extended the sparse group lasso to the high-dimensional multivariate response models with group structures in not only the predictors but also the responses. SCCS is a sequential method under the high dimensional multivariate regression model allowing complex group structures in both responses and predictors. It first selects coefficient block according to the canonical correlation coefficients and then selects nonzero coefficients in blocks by the EBIC \citep{chen2008extended} and often surpasses the MSGL in both accuracy and computation. Further, \citet{XIA2022108745} has studied a stepwise algorithm with the permutation importance measure, and \citet{luo2020feature} has discussed the principle of correlations. Sequential procedures also have been wild studied in other fields, such as decision making \citep{LIU2020105642}, fMRI data analysis \citep{FAN2020106341}, multidimensional time series \citep{9064715}, structural changes in time series \citep{kejriwal2020robust}, parameters estimation for radio equipment \citep{zaliskyi2021sequential}, etc.

Inspired by the related research and driven by empirical requirements, we focus on using the sequential procedure to detect the high-dimensional multiresponse models with group structures in both predictors and responses. In this paper, we propose a sequential method called sequential stepwise screening (SeSS) to detect the associations between predictor groups and response groups. Specifically, at each step, SeSS first selects a predictor group and a response group by calculating the correlation between predictor groups and the current residuals of response groups, then selects a predictor among the predictor group based on the correlation between each predictor and the current residual of the response group. The above strategy has two advantages: 1) It is used to identify the group structures and for the purpose of dimension reduction. 2) It is the key to control the computational time of the method rendering the computational complexity of the complex group structures roughly comparable with regularized regression modeling. We show that the proposed method enjoys the required theoretical properties, and show the effectiveness of this method in simulations and applications. In the empirical analysis, we apply the method to a GeneChip (Affymetrix) microarrays dataset and compare the results with the SCCS.

The rest of the paper is arranged as follows. In section 2, we introduce the models and describe SeSS in detail. In section 3, the main theoretical properties are provided. We present the simulation studies for the comparison of SeSS with the SCCS in section 4. In section 5, the real data is analyzed.

\section{Model and Methods}
In this section, we consider the sparse linear model in high-dimensional data with multiresponse and complex group structures, i.e., grouping responses and predictors according to prior information.
We allow the complex group structures, e.g., partial overlaps between groups and each response group may be related to multiple predictor groups. To handle these structures and obtain accurate estimates, we introduce the following framework and notations, which are mainly followed from \citet{luo2020feature}:
Set $q$ responses by $\mathcal{Y} = \left \{ y_1,\dots,y_q \right \}$, and the $j$th response group by $\mathcal{Y}_j$ $(j = 1,\dots, J)$. Similarly, set $p$ predictors by $\mathcal{X}=\left \{ x_1, \dots, x_p \right \}$ and the $k$th response group by $\mathcal{X}_k$ $(k = 1,\dots,K)$.  Groups can be overlapped, for example,  $\mathcal{Y}_1 = \{ y_1, y_2 \}$, $\mathcal{Y}_2 = \{ y_2, y_3, y_4 \}$, then $\mathcal{Y}_1 \cap \mathcal{Y}_2 = \{ y_2 \}$. We have $ \mathcal{Y} =\cup^{J}_{j=1}\mathcal{Y}_j $ and $ \mathcal{X} =\cup^{K}_{k=1}\mathcal{X}_k $. We use $ | \mathcal{X}_k |$ denote the number of variables in $\mathcal{X}_k$ and the same as $\mathcal{Y}_j$. Both $p$ and $q$ are much larger than $n$, but the size of $\mathcal{Y}_j$ and $\mathcal{X}_k$ are smaller than $n$.

Consider a sample of size $n$, let $Y$ be the $n \times q$ response matrix, and $X$ be the $n \times p$ predictor matrix. We suppose that each column of $X$ and $Y$ is standardized, i.e., the mean is zero and the variance is $n$. We denote $Y_j$ the matrix which consists the columns in $Y$ corresponding to the responses in group $\mathcal{Y}_j$, and $X_k$ the matrix which consists the columns in $X$ corresponding to the predictors in group $\mathcal{X}_k$. Note that both $\mathcal{Y}_j$, $\mathcal{X}_k$ can be overlapped, thus $Y$ and $X$ are generally not equal to $(Y_1,Y_2,\dots,Y_J)$ and $(X_1,X_2,\dots,X_K)$. For the notation simplicity, we redeclare $X$ and $Y$ as $Y = (Y_1,Y_2,\dots,Y_J)$ and $X = (X_1,\dots,X_K)$. We can write the model as follows:
\begin{equation*}
\big( Y_1, \dots, Y_J \big)
= \big( X_1, \dots, X_K \big)
\left( \begin{array}{ccc}
B_{11} & \dots, & B_{1J}\\
\dots, & \dots, & \dots,\\
B_{K1} & \dots, & B_{KJ}
\end{array}\right)
+ \big(\mathcal{E}_1, \dots,  \mathcal{E}_J \big),
\end{equation*}
where $\mathcal{E}_j$'s are random error matrices and $B_{kj}$ are the coefficient blocks.
\red{Each coefficient block \(B_{kj}\) describes the correlation of two groups, group \(X_k\) and group \(Y_j\).}
Denote the matrix consists of all the $B_{kj}$ blocks as $B$. We aim to select the relevant predictors for each group, that is, selecting nonzero elements of $B_{kj}$. To find the nonzero locations and estimate the coefficient matrix accurately, we propose a sequential procedure called SeSS. The procedure includes two parts. The first part focuses on finding the nonzero locations. Specifically, in this part, we select predictors for each response by selecting nonzero blocks in $B$; select nonzero rows in the selected block $B_{kj}$; select nonzero elements in the selected rows. In the second part, based on the obtained result, we conduct least-squared regression on the predictors to each response and screen the predictors for each response by imposing a threshold on the least-squared estimates.

Two measures are considered in the proposed method. One is the canonical correlation coefficients (CC), which is used in the first part that selects nonzero blocks in $B$ and then selects nonzero rows in $B_{kj}$. Let $\tilde Y_j$ be the residual matrix of $Y_j$ after conducting least-squared regression of certain predictors. We write the canonical correlation matrix between and \(X_k\) and \(\tilde Y_j\) as follows:
\[C_{kj} = \Sigma_k^{-1}\Xi_{kj}\Omega_j^{-1}\Xi_{kj},\]
where $\Sigma_k$, $\Omega_j$ and $\Xi_{kj}$ denote the variance matrices of $\mx_k$ and $\widetilde{\my}_j$ respectively and the covariance matrix between $\mx_k$ and $\widetilde{\my}_j$. It can be estimated by
\[\hat{C}_{k j}=\left(X_{k}^{\top} X_{k}\right)^{-1} X_{k}^{\top} \tilde{Y}_{j}\left(\tilde{Y}_{j}^{\top} \tilde{Y}_{j}\right)^{-1} \tilde{Y}_{j}^{\top} X_{k}.\]
Then we set the correlation measure as follows:
\[r(k, j)\triangleq r(X_k,\tilde Y_j) = tr(\hat{C}_{k j}).\]
The other measure is the extended Bayesian Information Criterion (EBIC) \citep{chen2008extended}, which is used as the stopping rule in selecting nonzero elements in the selected rows. Denote $r_{kj}$ as the number of nonzero elements in $B_{kj}$. We set $\zeta$ as a selected model with $m$ nonzero coefficient blocks. Let $\hat{B}^{(j)}(\zeta)$ be the least-square estimate of the $j$th column block, or, equivalently, the $(\hat{B}_{1j} \, \dots, \, \hat{B}_{Kj})^T$ and $\vert  B_{k_{l}j_{l}} \vert$ denote the number of total elements of $B_{k_{l}j_{l}}$. Then  we have
\begin{align*}
\text{EBIC}(\zeta) = & n\sum_{j=1}^{J}\ln\frac{1}{n} \big\| Y_j - X\hat{B}^{(j)}(\zeta)\big\|^2_F + \lambda_1\sum_{l=1}^{m}r_{k_lj_l}\ln n\\
& + 2\lambda_2 \gamma \left( \ln \binom{KJ}{m} + \ln \binom{\left | B_{k_lj_l} \right |}{r_{k_lj_l}} \right),
\end{align*}
where $\lambda_1$ and $\lambda_2$ are tuning parameters; $\| \cdot \|_F$ denotes the Frobenius norm; and $\gamma = 1 - \ln n/(2\ln p)$, following the setting by \citet{luo2020feature}. Based on the EBIC, we obtain $\tilde Y_j =  Y_j - X\hat{B}^{(j)}(\zeta)$. In the following, we show the selecting part of the proposed method.

\begin{center}
  Selection of SeSS
\end{center}
\vspace{-.5 cm}
Set $\zeta^* = \varnothing$.
\begin{itemize}
\item  \textit{Step 1.} Selection of coefficient blocks: Compute $r(k, j)$ for all the $k$ and $j$. If $r(k^*,{j^*})$ is the maximum, then the $B_{k^*j^*}$, $\left | \mathcal{X}_{k^*} \right | \times  \left | \mathcal{Y}_{j^*} \right |$ matrix, will be selected as the nonzero block in this step.
\end{itemize}
\begin{itemize}

\item \textit{Step 2.} Selection of coefficient rows: Let $B_{k^*j^*}  = (b_{k^*j^*}^1, \dots, b_{k^*j^*}^{|\mathcal{X}_{k^*}|})^T$.
% , where $(b_{k^*j^*}^a)^T$ is the $a$th row in $B_{k^*j^*}$.
Compute $r(X_{k^*}^{a},\tilde{Y}_{j^*})$ for all $a = 1,\dots,\left|\mathcal{X}_{k^*}\right| $
and choose the maximum, denoted as $r(X_{k^*}^{(a^*)},\tilde Y_{j^*})$ where $X_{k^*}^{a}$ is the $a$th column of $X_{k^*}$. In this step, a nonzero row will be selected,
% We note it as the $a^*$th row in $B_{k^*j^*}$,
denoted as $(b_{k^*j^*}^{a^*})^T$.
\end{itemize}
\begin{itemize}
\item \textit{Step 3.} Selection of nonzero elements:
% in $(b^{k^*j^*}_{a^*})^T$.
% Select the nonzero elements one at a time. Initially, l
Let $\mathcal{B}=\left \{ \beta_{a^*m} \right \}$ where $m = 1,\dots,|\mathcal{Y}_{j^*}|$ be the set of
%  all the elements of
row  $(b_{k^*j^*}^{a^*})^T$. Let $\zeta^*$ be the current model. Compute $ \text{EBIC}(\zeta^* \cup \beta_{a^*m})$ for all $\beta_{a^*m}$ in $\mathcal{B}$.
We note $\beta_{a^*\tilde{m}}$ as the minimum of $\{ \text{EBIC}(\zeta^*), \text{EBIC}(\zeta^* \cup \beta_{a^*m}), m = 1,\dots, |\mathcal{Y}_{j^*}|\}$. Update $\zeta^*$ to $\zeta^* \cup \beta_{a^*\tilde{m}}$ and simultaneously update $\mathcal{B}$ to $\mathcal{B} \setminus \beta_{a^*\tilde{m}}$. Repeat \textit{Step 3} until $\text{EBIC}(\zeta^* \cup \beta_{a^*\tilde{m}}) \geqslant  \text{EBIC}(\zeta^*)$ or $\mathcal{B} \setminus \beta_{a^*\tilde{m}}$ is empty. Then return to \textit{Step 2} and \textit{Step 1}.
\end{itemize}

\begin{figure}[!ht]
  \centering
  \includegraphics[scale=0.13]{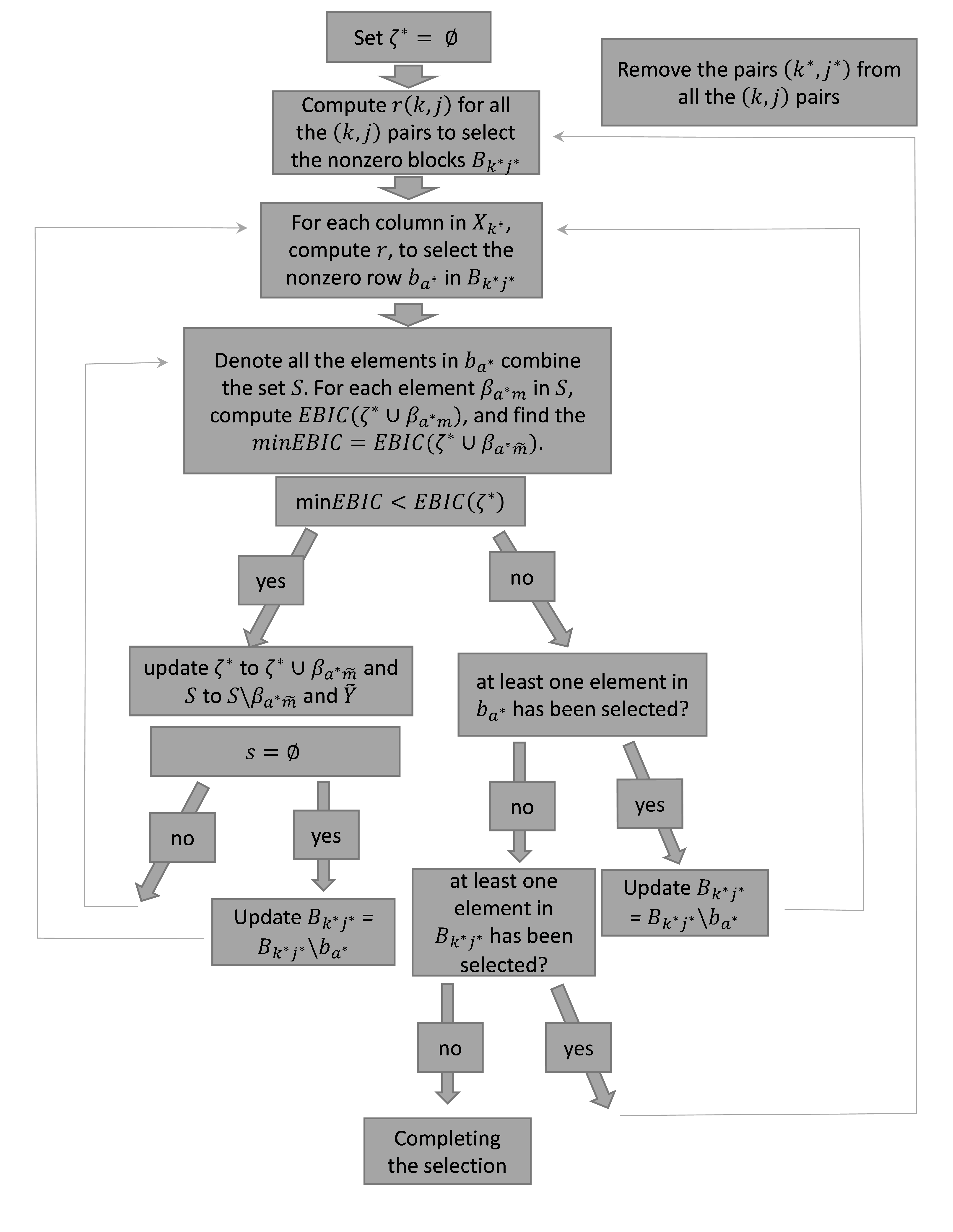}
  \caption{The selecting details of the SeSS algorithm.\label{fig 1}}
% 最后一步，completing the selection algorithm...something like that
\end{figure}

When returning to \textit{Step 2}, check whether at least one element in $(b^{k^*j^*}_{a^*})^T$ has been selected. If yes, in \textit{Step 2}, we update $B_{k^*j^*}$ as itself but without considering the row $(b_{k^*j^*}^{a^*})^T$. Else, which rarely happens, means that we do not select any element in row $(b_{k^*j^*}^{a^*})^T$. In this situation, we return directly to \textit{Step 1}. When returning to \textit{Step 1}, similar to above, check whether at least one element in $B_{k^*j^*}$ has been selected. If yes, the pairs $(k^*, j^*)$ will not be considered while computing $r(k, j)$ for all $(k, j)$ pairs. Else, the selection of the SeSS algorithm is completed.
We apply the ordinary least squares combined with a threshold $\rho$ to
obtain the estimate on the remaining entries of the coefficient matrix.
% \red{Then we apply the ordinary least squares to the selected nonzero entries, and at last conduct a fast screening of the estimations of the coefficient as if the absolute value of the estimate is smaller than a threshold $\rho$, shrink the estimate to zero.}
Figure~\ref{fig 1} exhibits the selection details of the proposed method.

In the algorithm, we obtain the model within the reduced feature space by EBIC and use the canonical correlation coefficients (CC) as the correlation measure. The former is shown with a small loss in the positive selection rate but tightly controls the false discovery rate in many applications \citep{chen2008extended}. For the CC, it is suitable for the linear model, compared with other measures such as distance correlation and Pearson's correlation. Distance correlation, for instance, measures the correlation including the nonlinear relationship among variables, which would lead to low power in a linear model detection \citep{li2012feature,kong2017interaction}.
\red{Though Pearson's correlation could measure the linear correlation between two variables, it can not describe the group correlation between two groups, while CC is often used and recommended for the data with group structures \citep{luo2020feature}.
}

% The former is inspired by the previous work \citep{luo2020feature},
% SeSS uses the canonical correlation coefficients (CC) as the correlation measure instead of the distance correlation (DC).
% The CC is often used and recommended for the data with group structures \citep{li2012feature,kong2017interaction}.
% We use the CC instead of the DC also because the DC measures the correlation including the nonlinear relationship among variables, which would lead to the low power of DC while in a linear model detection.

Compared with the SCCS, SeSS handles the sparse high-dimensional complex data more efficiently. For each selected group block, we propose adding a new step of selecting the most relevant predictor among the predictor group based on the correlation between each predictor and the current residual of the response group. This strategy has the following advantages in sparse high-dimensional models and computation: 1) avoids the exhaustivity of SCCS in each block and thus reduces the computational cost; 2) achieves high accuracy in detecting the extremely sparse models; 3) enjoys theoretical properties without extra requirement;
% Since the ultra-high dimensional data always lead to very sparse settings, the capacity of SeSS that can solve more sparse high-dimensional models is of much importance in practical data analysis with group structures.
4) SeSS's merits in the computation allow it to deal with the data with higher dimensions and more numerous group structures.

\section{Theoretical Properties}

In this section, we focus on the theoretical guarantee of the proposed method. We will show in the following that this method successfully recovers the true underlying sparse model with high probability. We first define some notations. Consider the following dimensional setting where $2\ln(pq) < n^{1/3 - \delta}$ that $0 < \delta < 1/3$ and $p_0 = O(n^{1/6})$ that each column of $B$ has at most $p_0$ nonzero elements. For each block $B_{kj}$ where $k = 1,\dots, K$ and $j = 1,\dots, J$, we assume the numbers of its column and row are bounded.

To study the asymptotic property of the proposed method, we first introduce a property of the canonical correlation measure established by \citet{luo2020feature}.
% , which discusses the properties of the canonical correlation measure.
Recall the notations of the canonical correlation matrix and its estimates, i.e., $C_{kj}(\zeta)$ and $\hat C_{kj}(\zeta)$ under model $\zeta$. Set $\tr(C_{k^*j^*}) = \max_{k, j} \{ \tr(C_{kj}(\zeta)) \}$. Then, under the following conditions, we have, as $n \rightarrow \infty$, uniformly for $\zeta$,
\begin{equation}\label{eq cannoical}
P\big( \tr(\hat C_{k^*j^*}(\zeta)) = \max_{k, j} \{ \tr( \hat C_{kj}(\zeta)) \} \big) \rightarrow 1.
\end{equation}
The proof of the above result can be found in the proof of Lemma~1 from \citet{luo2020feature}, so we omit it. We will display the required conditions of \eqref{eq cannoical}, A1 and A2, in the following.
Set $\mathcal{C}_0$ be the index set of nonzero columns of $B$ and set $\Psi_0$ be the index set of nonzero blocks of $B$. For the $l$th column of $B$ where $l \in \mathcal{C}_0$, we denote $s_{0l} = \{ i: \beta_{il} \neq 0 \}$ and $s^*_l$ be the nonzero index set of the response $y_l$ selected by the proposed method. Recall that $\Sigma_k$, $\Omega_j$ denote the variance matrices of $\mx_k$ and $\widetilde{\my}_j$. We use $X(s)$ to denote the subvector of $X$ with indices in $s$ and $\Sigma_{ss}$ to denote the related variance matrix. To state our theoretical results, we need the following conditions.

\begin{itemize}
\item[A1] The eigenvalues of $\Sigma_{k}$, $\Sigma_{s_{0l}s_{0l}}$, $\Omega_j$ are bounded from below and above for all $k,l, j$.
\item[A2] Let $\sigma(\cdot)$ be the standard deviation and let $t$ be a neighborhood of $0$. There exists a generic constant $C$ that $\max_{i, j}\{ \sigma(X_iX_j),\sigma(Y_iX_j),\sigma(X_iY_j) \} \leqslant C$ and $\max_{i, j}\big\{E\exp\{$\\$ t(X_iX_j - EX_iX_j)\}, E\exp\{ t(Y_iX_j - EY_iX_j)\}, E\exp\{ t(Y_iY_j - EY_iY_j)\}\big\} \leqslant C$.
\item[A3] Under model $\zeta$,
\[ \max_{(k, j) \notin \Psi_0} |\tr C_{kj}(\zeta)| < \max_{(k, j) \in \Psi_0} |\tr C_{kj}(\zeta)|, \]
and for $(k, j) \in \Psi_0$,
\[ \max_{l \notin \mathcal{C}_0} |\tr C^l_{kj}(\zeta)| < \max_{l \in \mathcal{C}_0} |\tr C^l_{kj}(\zeta)|. \]
\item[A4] Under model $\zeta$, for $l \in \Psi_0$ and $|\zeta| < |\Psi_0|$,
\[  \max_{i \in s^c_{0l}} \ln \dfrac{ (I - H(\zeta)) XB^{(j)}(\zeta) }{(I - H(\zeta)) XB^{(j)}(\zeta \cup i)}  \leqslant   \ln n\lambda_1/n,\]
where $H(\zeta) =  X(\zeta)[X^\t(\zeta)X(\zeta)]^{-1} X^\t(\zeta)$.
\item[A5] As $n \rightarrow \infty$,
\[\sqrt{n} \min_{(k, j) \in \Psi_0} \min_{l \in \mathcal{Y}_k, i \in s_{0l} \cap \mathcal{X}_j} |\beta_{il}|/\sqrt{p_0 \ln p} \rightarrow \infty. \]
\end{itemize}

A1 - A2 are used for proving \eqref{eq cannoical}. A1 is a regular condition that assumes the eigenvalues of the true covariance matrix are bounded. A2 is raised from \citet{fill1983convergence}. These two conditions are used to provide the error bound between the associated sample covariance and the true covariance. A3 - A4 are used to prove that the proposed method is selection consistent. A5 states that the relative portion of a variation of the irrelevant predictor is bounded by the tuning parameter. These conditions are all followed from \citet{luo2020feature}.
Then we have the following result of the selection consistency of SeSS. The proof of Theorem~\ref{thm} is given in Appendix.

\begin{thm}\label{thm}
Suppose A1 - A5 hold. The SeSS is selection consistent, i.e., with probability tends to $1$ as $n \rightarrow \infty$, we have $s^*_l = s_{0l}$ for $l \in \mathcal{C}_0$ and $s^*_l = \emptyset$ for $l \in \mathcal{C}^c_0$.
\end{thm}

\section{Simulations}
\red{In this section, we will demonstrate the performance of the SeSS and compare it with the SCCS and the other two methods: group LASSO \citep{yang2015fast} and elastic net \citep{friedman2010regularization}. We also test the performance of the backward selection \citep{tsagris2018feature} and forward-backward selection \citet{borboudakis2019forward}. They both suffer from the complexity of group structures and do not perform well, thus their results are not presented.
We consider two cases of the group structure, an equal size case and an unequal size case. We also consider four settings of sparsity, i.e., $90\%$, $95\%$, $70\%$, $50\%$} of elements of the coefficient matrix equals zero. During the simulations, we consider two dimensions, $n=150, q=200, p=200$ or $n=150, q=200, p=400$. In both dimensions, we consider both group structures in predictors and responses. Simulation settings are presented in detail in the following.

We generate each row of $X$ independently from a multivariate normal distribution $N(0,\Sigma)$, where $\Sigma_{ij} = 0.5^{|i-j|}$. Elements of the error matrix $\mathcal{E}$ are generated from $N(0,\sigma^2)$ independently and $\sigma^2$ depends on the variance of $XB$. We denote $z_l$ be the $l$th column of $XB$ and $V_1 = \sum^q_{l=1}var(z_l)$. Then we set $\sigma^2 = \frac{V_1}{5q}$. We consider the diagonal block setting, that is, for the coefficient blocks $B_{kj}$ in $200 \times 200$ or $400 \times 200$ coefficient matrix, we have $B_{kj} = 0$, $k \neq j$. For diagonal blocks, we first set the values of coefficient matrix independently generated from a uniform distribution on $[-5,-1] \cup [1,5]$, then randomly select $90\%$, $95\%$, $70\%$, $50\%$ of them and set to zero. For the group structures, we consider two cases:
\begin{itemize}
  \item[] \text{Equal-size case:} The group size of responses and predictors are both equal $20$.
  \item[] \text{Unequal-size case:} The group size of responses and predictors are randomly set as $20$ or $30$.
\end{itemize}
Each group consists of the variables in order, for example, $\mathcal{Y}_j = \{ y_{ 20(j-1)+1}, \dots,$\\$ y_{20j} \}$ and $\mathcal{X}_k = \{x_{ 20(k-1)+1}, \dots,x_{20k} \} $, where $j,k = 1,\dots,20$. \red{For the proposed method, the threshold $rho$ is set inspired from \citet{guo2016spline}, that is, we set $\rho = \hat \sigma \sqrt{2\log p}$ where $\hat \sigma$ is the standard error of the estimated coefficients and is adjusted by small magnitude. We also use this threshold in Real example.}
We use \red{seven} measures to show the performance of the \red{four} methods: $l_1$-norm error norm, $l_2$-norm error, positive discovery rate (PDR) and false discovery rate (FDR), discovery rate (DR), \red{blocks discovery rate (BDR)} and computational cost. $l_1$-norm error denotes $\|\hat B - B\|_1$ and $l_2$-norm error denotes the Frobenius norm error $\|\hat B - B\|_F$. For the PDR and FDR, we have,  with $\theta_{il} = I\{ \beta_{il} \neq 0\}$:
\[ \text{PDR} = \frac{ \# \{ \hat{\theta}_{il} = \theta_{il} = 1  \} }{ \# \{ \theta_{il} = 1 \} },
     \text{FDR} = \frac{ \# \{ \hat{\theta}_{il} = 1 ,\theta_{il} = 0  \} }{ \# \{ \hat{\theta}_{il} = 1\} }.\]
Both PDR and FDR are one-sided since we can either select most elements to get a high PDR or simply choose a few elements in $B$ for a low FDR. So we also introduce $\text{DR} = \text{PDR}+1-\text{FDR}$ as a joint measure. \red{Similar to DR, BDR is a measure describing the accuracy of the selected blocks while DR is a measure describing the accuracy of the selected entries.} The computational costs are performed on a standard laptop computer with a 2.50 GHz Intel Core i5-7200U processor.

%\newpage

% \usepackage{graphicx}

\begin{table}[!htbp]
\centering
\caption{\red{The average PDR, FDR, DR, BDR, $l_1$-norm error, $l_2$-norm error, and the computation time over 100 times simulation studies in the cases of 0.9, 0.95 sparsity(numbers in parentheses are standard deviations).}}
\label{tab:table1}
\resizebox{\textwidth}{!}{%
\begin{tabular}{ccccccccc}
\hline
Method      & Sparsity          & PDR           & FDR           & DR            & BDR           & L1                & L2                & time            \\ \hline
            & Equal-size        &               & n=150         & q=200         & p=200         &                   &                   &                 \\ \hline
SeSS        & 0.9               & 0.913(0.014) & 0.009(0.005) & 1.903(0.017) & 1.866(0.141) & 134.049(11.213)  & 131.426(24.903)  & 44.414(6.399)  \\
SCCS        & 0.9               & 0.939(0.008) & 0.059(0.008) & 1.880(0.014) & 1.848(0.107) & 122.563(9.765)   & 93.965(18.853)   & 217.514(5.803) \\
group Lasso      & 0.9          & 1(0)          & 0.989(0.001)  & 1.011(0.001)  & 1.097(0.166) & 1599.585(32.888)  & 3088.827(148.376) & 49.384(0.735)   \\
elastic net & 0.9               & 1(0)          & 0.978(0.001)  & 1.022(0.001)  & 1.100(0)     & 1521.571(33.471)  & 2950.001(146.11)  & 34.743(1.135)   \\ \hline
SeSS        & 0.95              & 0.967(0.012) & 0.002(0.002) & 1.966(0.013) & 1.845(0.163) & 35.636(4.872)   & 21.333(8.876)   & 28.816(4.097)  \\
SCCS        & 0.95              & 0.934(0.018) & 0.073(0.032) & 1.862(0.030) & 1.097(0.125) & 52.871(6.083)   & 42.555(13.802)   & 116.530(8.343) \\
group Lasso      & 0.95         & 1(0)          & 0.995(0.001)  & 1.005(0.001)  & 1.134(0.149) & 1230.227(22.781)  & 1440.917(43.205)  & 50.635(0.886)   \\
elastic net & 0.95              & 1(0)          & 0.989(0.002)  & 1.011(0.002)  & 1.100(0)     & 1155.623(25.233)  & 1373.527(40.095)  & 39.995(0.945)   \\ \hline
            & Unequal-size      &               & n=150         & q=200         & p=200         &                   &                   &                 \\ \hline
SeSS        & 0.9               & 0.872(0.012) & 0.011(0.006) & 1.861 (0.015) & 1.882(0.138) & 262.216(15.239) & 498.435(63.668) & 42.532(6.224)  \\
SCCS        & 0.9               & 0.925(0.018) & 0.083(0.047) & 1.842 (0.051) & 1.842(0.11)  & 191.298(26.039) & 150.094(36.369) & 507.775(8.625) \\
group Lasso      & 0.9          & 1(0)          & 0.987(0.001)  & 1.013(0.001)  & 1.115(0.185) & 1936.406(27.026)  & 4095.843(126.883) & 56.394(1.315)   \\
elastic net & 0.9               & 1(0)          & 0.971(0.001)  & 1.029(0.001)  & 1.100(0)     & 1816.066(24.892)  & 3903.617(125.112) & 33.143(0.463)   \\ \hline
SeSS        & 0.95              & 0.934(0.013) & 0.002(0.003) & 1.931(0.012) & 1.848(0.117) & 66.246(7.532)   & 59.924(17.554)   & 34.478(5.196)  \\
SCCS        & 0.95              & 0.954(0.015) & 0.068(0.018) & 1.886(0.029) & 1.868(0.124) & 63.128(7.084)   & 40.550(11.482)   & 270.149(6.861) \\
group Lasso      & 0.95         & 1(0)          & 0.993(0.001)  & 1.007(0.001)  & 1.124(0.112) & 1365.569(17.51)   & 1962.492(62.723)  & 56.3(1.463)     \\
elastic net & 0.95              & 1(0)          & 0.986(0.001)  & 1.014(0.001)  & 1.100(0)     & 1245.935(22.097)  & 1850.772(63.71)   & 37.72(0.738)    \\ \hline
            & Equal-size        &               & n=150         & q=200         & p=400         &                   &                   &                 \\ \hline
SeSS        & 0.9               & 0.908(0.014)  & 0.012(0.005)  & 1.896(0.018)  & 1.905(0.08)  & 138.489(12.357)   & 140.876(29.172)   & 56.018(9.465)   \\
SCCS        & 0.9               & 0.959(0.009)  & 0.080(0.015)  & 1.879(0.015)  & 1.927(0.209) & 111.935(7.506)    & 63.230(10.356)    & 233.501(4.446)  \\
group Lasso      & 0.9          & 1(0)          & 0.994(0.002)  & 1.006(0.002)  & 1.157(0.100) & 1586.44(31.077)   & 3122.184(150.279) & 73.859(1.252)   \\
elastic net & 0.9               & 1(0)          & 0.979(0.001)  & 1.021(0.001)  & 1.100(0)     & 1427.115(30.759)  & 2983.029(149.058) & 54.341(0.966)   \\ \hline
SeSS        & 0.95              & 0.958(0.008)  & 0.003(0.004)  & 1.956(0.011)  & 1.929(0.178) & 41.069(4.699)     & 34.854(10.621)    & 36.246(4.763)   \\
SCCS        & 0.95              & 0.959(0.009)  & 0.080(0.015)  & 1.879(0.015)  & 1.938(0.147) & 111.935(7.506)    & 63.230(10.356)    & 127.970(4.080)  \\
group Lasso      & 0.95         & 1(0)          & 0.997(0.001)  & 1.003(0.001)  & 1.129(0.1)   & 1191.112(15.741)  & 1445.955(45.749)  & 73.641(1.152)   \\
elastic net & 0.95              & 1(0)          & 0.989(0.001)  & 1.011(0.001)  & 1.100(0)     & 994.38(20.28)     & 1369.649(43.083)  & 62.657(1.25)    \\ \hline
            & Unequal-size &    & n=150         & q=200         & p=400         &                   &                   &                 \\ \hline
SeSS        & 0.9               & 0.894(0.010)  & 0.021(0.006)  & 1.873(0.014)  & 1.821(0.189) & 232.398(22.567)   & 361.499(79.828)   & 59.067(5.033)   \\
SCCS        & 0.9               & 0.919(0.012)  & 0.038(0.010)  & 1.881(0.014)  & 1.898(0.163) & 180.875(11.228)   & 150.691(29.439)   & 500.800(7.355)  \\
group Lasso      & 0.9          & 1(0)          & 0.992(0)      & 1.008(0)      & 1.125(0.157) & 1932.355(25.273)  & 4155.343(132.346) & 84.788(1.843)   \\
elastic net & 0.9               & 1(0.001)      & 0.973(0.001)  & 1.027(0.001)  & 1.100(0)     & 1732.345(28.41)   & 3958.264(130.069) & 54.365(0.656)   \\ \hline
SeSS        & 0.95              & 0.941(0.012)  & 0.006(0.006)  & 1.935(0.017)  & 1.886(0.123) & 63.996(8.525)     & 57.772(19.750)    & 37.254(2.306)   \\
SCCS        & 0.95              & 0.910(0.015)  & 0.025(0.008)  & 1.885(0.016)  & 1.825(0.237) & 79.403(4.414)     & 78.503(10.107)    & 276.498(8.820)  \\
group Lasso      & 0.95         & 1(0)          & 0.996(0)      & 1.004(0)      & 1.111(0.14)  & 1341.372(17.375)  & 1984.445(67.73)   & 82.678(1.844)   \\
elastic net & 0.95              & 1(0)          & 0.986(0.001)  & 1.014(0.001)  & 1.100(0)     & 1112.189(24.504)  & 1859.54(66.452)   & 59.645(0.654)   \\ \hline
\end{tabular}%
}
\end{table}

% \newpage
\begin{table}[!htbp]
% \usepackage{graphicx}
\centering
\caption{\red{The average PDR, FDR, DR, $l_1$-norm error, and $l_2$-norm error and the computation time over 100 times simulation studies in the cases of 0.5, 0.7 sparsity(numbers in parentheses are standard deviations).}}
\label{tab:table2}
\resizebox{\textwidth}{!}{%
\begin{tabular}{cccccccc}
\hline
Method & Sparsity     & PDR          & FDR          & DR           & L1                & L2                & time           \\ \hline
       & Equal-size   &              & n=150        & q=200        & p=200             &                   &                \\ \hline
SeSS   & 0.7          & 0.787(0.009) & 0.097(0.008) & 1.69(0.015)  & 1239.473(37.873)  & 1692.364(94.373)  & 21.026(0.456)  \\
SCCS   & 0.7          & 0.91(0.011)  & 0.047(0.01)  & 1.864(0.016) & 590.978(29.409)   & 516.614(50.189)   & 167.753(3.565) \\
SeSS   & 0.5          & 0.611(0.005) & 0.334(0.009) & 1.277(0.011) & 3660.833(57.515)  & 6853.23(227.779)  & 39.43(0.346)   \\
SCCS   & 0.5          & 0.835(0.011) & 0.028(0.004) & 1.807(0.011) & 1416.775(55.531)  & 1718.264(141.828) & 227.277(3.072) \\ \hline
       & Unequal-size &              & n=150        & q=200        & p=200             &                   &                \\ \hline
SeSS   & 0.7          & 0.627(0.006) & 0.353(0.013) & 1.274(0.017) & 2945.466(67.337)  & 5018.946(231.614) & 54.179(0.661)  \\
SCCS   & 0.7          & 0.795(0.017) & 0.022(0.004) & 1.773(0.019) & 1124.532(54.125)  & 1503.029(149.8)   & 337.738(6.832) \\
SeSS   & 0.5          & 0.416(0.008) & 0.51(0.017)  & 0.906(0.022) & 6914.002(219.93)  & 16193.394(750.4)  & 69.153(2.666)  \\
SCCS   & 0.5          & 0.453(0.169) & 0.019(0.004) & 1.434(0.167) & 4523.972(1235.173) & 11940.482(5496.638) & 298.133(99.075) \\ \hline
       & Equal-size   &              & n=150        & q=200        & p=400             &                   &                \\ \hline
SeSS   & 0.7          & 0.786(0.008) & 0.073(0.006) & 1.713(0.012) & 1208.827(36.276)  & 1627.983(95.574)  & 28.168(0.986)  \\
SCCS   & 0.7          & 0.915(0.008) & 0.027(0.007) & 1.887(0.011) & 536.351(17.082)   & 448.209(37.127)   & 164.131(2.1)   \\
SeSS   & 0.5          & 0.595(0.011) & 0.335(0.011) & 1.259(0.02)  & 3747.004(80.347)  & 7097.174(284.317) & 58.544(1.868)  \\
SCCS   & 0.5          & 0.844(0.011) & 0.016(0.002) & 1.827(0.011) & 1293.958(54.144)  & 1493.908(143.688) & 229.054(2.569) \\ \hline
       & Unequal-size &              & n=150        & q=200        & p=400             &                   &                \\ \hline
SeSS   & 0.7          & 0.609(0.01)  & 0.365(0.015) & 1.244(0.022) & 3127.801(96.576)  & 5428.155(305.734) & 80.475(1.763)  \\
SCCS   & 0.7          & 0.796(0.015) & 0.015(0.002) & 1.782(0.016) & 1092.801(52.359)  & 1450.153(147.862) & 335.775(6.808) \\
SeSS   & 0.5          & 0.395(0.005) & 0.549(0.008) & 0.846(0.01)  & 7378.296(104.048) & 17083.66(493.544) & 106.178(2.14)  \\
SCCS & 0.5 & 0.234(0.186) & 0.014(0.007) & 1.22(0.185)  & 6063.39(1342.351)  & 18839.385(5953.104) & 167.59(103.51)  \\ \hline
\end{tabular}%
}
\end{table}

For each situation, we simulate 100 times. \red{As one can see from Table~\ref{tab:table1}, we can find that SeSS performs better than the SCCS in FDR, DR, and computational time in all the cases. And the other two methods such as group LASSO and elastic net do not have a good performance because they both have a high FDR, which means they select many wrong nonzero entries. From the BDR metrics, we can see that both SeSS and SCCS almost select features in the correct blocks while the other two cannot. We only compare the performance of SeSS and SCCS in the cases of $0.3$ and $0.5$ sparsity to show that SeSS and SCCS are not very suitable for the data with low sparsity. Table~\ref{tab:table2} describes the simulation in the low sparsity cases of 0.5 and 0.7 sparsity. We can see that both SeSS and SCCS do not perform well in these cases. However, in all the cases SCCS has a higher DR than SeSS and lower L1 and L2. This shows that SCCS has the capability to deal with the data with a wide range of sparsity.}

On average, the computational time used for SeSS is generally $1/10$ to $1/6$ that of SCCS. In addition, we can find that in both the equal case and the unequal case, when the sparsity is 0.9, the PDR of SeSS is slightly lower than SCCS but the joint measure DR is still comparable. When the sparsity is 0.95, SeSS, in general, outperforms SCCS. Both tables show that the SeSS algorithm is efficient in computation and always achieves a low FDR, which are two merits in terms of the quality of feature selection.

\section{Real example}
In this section, we apply the proposed method to a GeneChip (Affymetrix) microarrays dataset, which was used in \citet{wille2004sparse} to reverse engineer genetic regulatory networks by using GGM (Graphical Gaussian Modeling). The dataset is about 118 GeneChip (Affymetrix) microarrays for the expressions of 39 genes, which are in the isoprenoid pathways in \textit{Arabidopsis thaliano} and 21 genes are in the \textit{mevalonatepatty} and 18 in the \textit{nonmevalonatepatty} pathway. 795 additional genes from 56 downstream metabolic pathways are also consisted in the dataset. In order to formulate the genetic regulatory network, we use a conditional Gaussian graphical model with group structures: $Y = X \mathcal{B} + \mathcal{E} $ with $(n,p,q) = (118,795,39)$.
$Y$ denotes the observation of the expression levels of genes in the isoprenoid pathways and $Y$ has a group structure $Y = (Y_1, Y_2)$. $X$ denotes the observation of the expression levels of genes in the downstream metabolic pathways which is divided into 56 groups as $X = (X_1,\dots, X_{56})$.

We respectively use SCCS, SeSS, group LASSO, and elastic net to estimate the coefficient matrix $\mathcal{B}$.
\red{Because the real $\mathcal{B}$ is unknown in empirical analysis, we do not use either the PDR or the FDR or the DR as the performance measure. Instead, we use the mean-squared prediction error (MSPE) to describe the performance of four methods.}
In each test, we randomly choose $n_0 = 70$ or $n_0 = 100$ microarrays of the 118 microarrays as the training sample and the remaining 48 and 18 microarrays as the testing sample. We then use the training sample to select the nonzero entries of $B$ and estimate it by the ordinary least squares. The prediction error is computed using the testing sample. We repeat the test 100 times and the prediction errors are averaged and reported in Table~\ref{tab:table3}. We also show the number of the nonzero entries (NNE) selected by the algorithm and the computational time of SCCS and SeSS respectively. The mean-squared error (MSE) computed by the training sample is shown as a baseline of the MSPE.
From the result reported in Table~\ref{tab:table3} where $n_0$ equals 100, we can see that the mean MSPE of SeSS is 6.9\% less than that of SCCS. \red{The group Lasso and elastic net both have lower MSE and MSPE than SeSS and SCCS, however, both methods select a great number of nonzero entries.} There is no order of magnitude difference between the NNE of SeSS and SCCS. The computational merits of SeSS surpass the SCCS, e.g., the SCCS requires 195 sec while SeSS requires 29 sec, which is 15\% of the former. \red{When the $n_0$ equals 70, the group Lasso and elastic net perform much worse than the former case with higher MSE and MSPE than SeSS and SCCS.} The mean MSPE of SeSS is 2\% less than that of SCCS. At this time, there is a significant difference in the NNE compared with the former. SeSS is computationally attractive compared with SCCS in that the time SeSS used is only 10.9\% of the time SCCS used.

\begin{table}[!htbp]\small%
\centering
\caption{\red{The average MSE, MSPE, NNE, and computational time (numbers in parentheses) are standard deviations.}}
\label{tab:table3}
\begin{tabular}{ccccc}
\hline
            & MSE          & MSPE         & NNE               & Time            \\ \hline
$n_0 = 100$ &              &              &                   &                 \\ \hline
SeSS        & 0.419(0.089) & 0.613(0.146) & 144.950(11.452)   & 29.431(14.546)  \\
SCCS        & 0.415(0.014) & 0.659(0.212) & 112.820(5.528)    & 195.239(25.938) \\
group Lasso      & 0.275(0.026) & 0.579(0.171) & 3208.512(216.136) & 46.054(2.269)   \\
elastic net & 0.278(0.020) & 0.520(0.147) & 1194.323(77.400)  & 5.791(0.179)    \\ \hline
$n_0 = 70$  &              &              &                   &                 \\ \hline
SeSS        & 0.295(0.072) & 0.674(0.137) & 211.03(60.458)    & 39.201(13.112)  \\
SCCS        & 0.228(0.013) & 0.688(0.126) & 276.2(42.367)     & 358.75(48.623)  \\
group Lasso      & 0.402(0.072) & 0.655(0.152) & 2560.473(288.380) & 36.925(3.368)   \\
elastic net & 0.312(0.037) & 0.620(0.162) & 839.217(49.376)   & 4.814(0.205)    \\ \hline
\end{tabular}%
\end{table}

\section{Summary}
For the high-dimensional multiresponse models with complex group structures, the proposed method efficiently detects the related responses and predictors.
Inspired by SCCS, SeSS is based on the canonical correlation and extended Bayesian Information Criterion. We innovatively propose a three-steps sequential screening algorithm. Simulations and real example analysis both show that SeSS has a better performance than SCCS on these data of particular group structures.

In future research, two possible improvements of SeSS are worth studying. One is the replacement of the canonical correlation and the extended Bayesian Information Criterion. The other is the trade-off between computational cost and accuracy. A suggestion is to address the accuracy problem induced by the greedy algorithm, we can try selecting an entry in the nonzero row and then changing the target nonzero row.
\red{Further, another attractive idea of dealing with the models with complex group structures is using deep learning methods. Some researchers have studied relevant research on feature selection or regression by using deep learning methods \citep{niu2020developing,qiu2014ensemble}. It would be worthwhile to study the vast amounts of data with deep neural networks and other deep learning strategies.}
The study of high-dimensional multiresponse models with complex group structures is of great significance, and more relevant studies are desired.

\section*{Acknowledgement}
This work was supported by the National Natural Science Foundation of China (Grant No. 12001557); the Youth Talent Development Support Program (QYP202104), the Emerging Interdisciplinary Project, and the Disciplinary Funding of Central University of Finance and Economics.

\section*{Compliance with ethical standards}
The authors declare no potential conflict of interests. The authors declare no research involving human participants and/or animals. The authors declare informed consent.

\appendix
\section*{Appendix}
\red{Before we give the proof of Theorem~1, we first give the following result.}
\begin{lem}\label{lem}
Under the same conditions of Theorem~\ref{thm}. As $n \rightarrow \infty$, the EBIC is selection consistent in the following sense:
\[ P(\min_{|\zeta| < |\zeta^*|} EBIC(\zeta) > EBIC(\zeta^*) ) \rightarrow 1.  \]
\end{lem}
\red{The proof of Lemma~1 is given in  \citep{luo2013extended}, so we omit the proof here.}

\begin{proof}[Proof of Theorem~1]
Recall that
\begin{equation*}
P\big( \tr(\hat C_{k^*j^*}(\zeta)) = \max_{k, j} \{ \tr( \hat C_{kj}(\zeta)) \} \big) \rightarrow 1.
\end{equation*}
Above result is used for Step 1. In Step 2, consider each nonzero block, $k^*$ and $j^*$ and set $\tr(C^{a^*}_{k^*j^*}) = \max_{a} \{ \tr(C_{k^*j^*}(\zeta)) \}$ \red{where $a$ denotes the $a$th row of the nonzero block.} We need to prove that as $n \rightarrow \infty$, uniformly for $\zeta$,
\begin{equation*}
P\big( \tr(\hat C^{a^*}_{k^*j^*}(\zeta)) = \max_{a} \{ \tr( \hat C^a_{k^*j^*}(\zeta)) \} \big) \rightarrow 1.
\end{equation*}
With a little abuse of notation, it suffices to show that for $j$, $k$, $\zeta$ and $a$, we have
\[ \tr(\hat C_{kj}^a) = \tr(C^a_{kj}(\zeta)(1+ o_p(1))).\]
Based on the proof of Lemma~1 of \citet{luo2020feature}, we have that,
\[ \tr(\hat C_{kj}) = \tr(C_{kj}(\zeta)(1+ o_p(1))),\]
leading to the result directly. Thus, we have that when $(k, j) \notin \Psi_0$, $\hat B^*_{kj} = 0$. Then we consider $(k, j) \in \Psi_0$. Given $(k, j)$, set $\zeta^*$ to be the current model, and set $\zeta^*_{-1}$, $\zeta^*_{+1}$ be the model obtained by the last step and the next step respectively. Based on the nature of the sequential procedure, we have $\zeta^*_{-1} \subset \zeta^* \subset \zeta^*_{+1}$. To obtain the true model, it suffices to prove the following two cases: Part 1. The algorithm converges \red{when the number of selected nonzero blocks equals to \(| \Psi_0 |\)}; Part 2. The irrelevant predictor would not be chosen by any step.
% First, we show that when $|\zeta^*{-1}| \subset $
First, assume that the size of $\widetilde{\zeta}$ equals that of the true model.
If $\zeta^* = \widetilde \zeta$, \red{to prove the first part is sufficient to proving that}
\[ \min\{\ebic(\zeta^*_{-1}) , \ebic(\zeta^*_{+1})\} > \ebic(\zeta^*). \]
At the current step, only the current nonzero coefficient block is relevant, that is
\[\text{EBIC}_{\my_j}(\zeta^*) = n \ln\frac{1}{n} \big\| Y_j - X\hat B^{(j)}(\zeta^*)\big\|^2_2 + \lambda_1 r \ln n + \lambda_2\cdot 2 \gamma \ln \binom{|B_{kj}|}{r},\]
where $r$ denotes the number of nonzero entries of the current mode of the $kj$th coefficient block.
We need to prove that for each \red{$l$ column that belongs to the nonzero blocks}, the zero elements of this column, i.e., $i \in s_{0l}^c$, cannot be chosen by EBIC.
Further, let $D_+ = \text{EBIC}_{\my_j}(\zeta^*) - \text{EBIC}_{\my_j}(\zeta^*_{+1})$, $D_- = \text{EBIC}_{\my_j}(\zeta^*_{-1}) - \text{EBIC}_{\my_j}(\zeta^*)$. It suffices to show that if $\zeta^*$ indicates the true model, then we have $D_+ <0$ and $D_- > 0$.

For proving $D_- > 0$, as shown in Lemma~\ref{lem}, we can conclude $D_- > 0$ based on the property of the EBIC. For proving $D_+ <0$, similar as the proof of Theorem~1 of \citet{luo2020feature}, we have
\begin{align}\label{eq d}
D_+  = n \ln \dfrac{\big\| Y_j - X\hat B^{(j)}(\zeta^*)\big\|^2_2 }{\big\| Y_j - X\hat B^{(j)}(\zeta^*_{+1})\big\|^2_2 } - \lambda_1 \ln n + 2 \lambda_2 \Big[\ln \binom{|B_{kj}|}{r} - \ln \binom{|B_{kj}|}{r+1}\Big].
\end{align}
Following the sparse assumption and setting $\widetilde{\zeta}_{kj}$ be the true model in the current blocks, we have that $|B_{kj}| \gg |\widetilde{\zeta}_{kj}|$ and thus $|B_{kj}| - r > r +1$. We have
\[ \ln \binom{|B_{kj}|}{r} - \ln \binom{|B_{kj}|}{r+1} =
\ln \dfrac{|B_{kj}|!(r+1)! (|B_{kj}| - r - 1)! }{|B_{kj}|!  r! (|B_{kj}| - r)! }
= \ln \dfrac{r+1}{|B_{kj}| - r} < 0. \]
% We have the last term on the right hand of the above equation less than zero.
Further, we have
\begin{align*}
 \big\| Y_j - X\hat B^{(j)}(\zeta^*)\big\|^2_2 & = \|(I - H(\zeta^*)) Y_j\|^2_2\\
& = \|(I - H(\zeta^*)) XB^{(j)}(\zeta^*)\|^2_2 (1+ o(1)),
\end{align*}
where $H(\zeta^*) =  X(\zeta^*)[X^\t(\zeta^*)X(\zeta^*)]^{-1} X^\t(\zeta^*)$. The last equality holds following the proof of Theorem~3.1 of \citet{luo2014sequential}. The first term on the right hand of \eqref{eq d} thus can be written as
\begin{align*}
n \ln \dfrac{\big\| Y_j - X\hat B^{(j)}(\zeta^*)\big\|^2_2 }{\big\| Y_j - X\hat B^{(j)}(\zeta^*_{+1})\big\|^2_2 } & = n \ln \dfrac{ (I - H(\zeta^*)) XB^{(j)}(\zeta^*) }{(I - H(\zeta^*)) XB^{(j)}(\zeta^*_{+1})}.
\end{align*}
Based on \red{A4}, the above term is less $\lambda_1 \ln n$, indicating $D_+ <0$, Part 1 is completed. Following the arguments as the proof of $D_+ <0$, we have that when for each irrelevant predictor $i$, EBIC$(\zeta^*)$ - EBIC$(\zeta^* \cup i) < 0$, thus $i$ cannot be chosen, completing the proof.
\end{proof}

\bibliography{reference}

\end{document}